\documentclass{article}
\usepackage[T1]{fontenc}
\usepackage{graphicx}
\usepackage[pagebackref=true]{hyperref}
\usepackage{amsmath}
\usepackage{amsfonts}
\usepackage{amssymb}
\usepackage{bbold} 
\usepackage{url}
\usepackage{fancyhdr}
\usepackage{booktabs}
\usepackage{color}
\usepackage[ruled, linesnumbered]{algorithm2e}
\usepackage{ulem} 
\usepackage{cleveref} 
\usepackage{mathtools} 
\usepackage[a4paper, total={5.9in, 9.3in}]{geometry}

\newcommand{\V}{\mathcal{V}}
\newcommand{\E}{\mathcal{E}}
\newcommand{\G}{\mathcal{G}}
\newcommand{\T}{\mathcal{T}}

\newcommand\PG[2]{\textcolor{magenta}{#1}}

\begin{document}
\title{\textbf{Temporal network compression via \\ network hashing}}

\author{R\'{e}mi Vaudaine$^1$ \and Pierre Borgnat$^{2}$\and Paulo Goncalves$^{1}$\and R\'{e}mi Gribonval$^{1}$\and M\'arton Karsai$^{3,4,*}$}

\date{%
\small \emph{
    $^1$ Univ Lyon, EnsL, UCBL, CNRS, Inria, LIP, F-69342, Lyon Cedex 07, France \\%
    $^2$ CNRS, Univ de Lyon, ENS de Lyon, Laboratoire de Physique, F-69342 Lyon, France\\%
    $^3$ Department of Network and Data Science, Central European University, 1100 Vienna, Austria\\%
    $^4$ Rényi Institute of Mathematics, 1053 Budapest, Hungary\\%
    $^*$Corresponding author: mkarsai@ceu.edu}
}

\maketitle              
\begin{abstract}

Pairwise temporal interactions between entities can be represented as temporal networks, which code the propagation of processes such as epidemic spreading or information cascades, evolving on top of them. The largest outcome of these processes is directly linked to the structure of the underlying network. Indeed, a node of a network at given time cannot affect more nodes in the future than it can reach via time-respecting paths. This set of nodes reachable from a source defines an out-component, which identification is costly. In this paper, we propose an efficient matrix algorithm to tackle this issue and show that it outperforms other state-of-the-art methods. Secondly, we propose a hashing framework to coarsen large temporal networks into smaller proxies on which out-components are easier to estimate, and then recombined  to obtain the initial components. Our graph hashing solution has implications in privacy respecting representation of temporal networks.
\\\\
\textbf{keywords:} Temporal networks, out-component calculation, streaming matrix algorithms, graph hashing
\end{abstract}

\section{Introduction}
While temporal networks represent the sequence of time-evolving interactions between entities, they also code the connected structure that lays behind many dynamical processes like the spreading of an epidemic or an information cascade or the collective adoption of behavioural norms or products.
In static networks, connectivity is conventionally defined between two nodes if they are connected via a direct edge, or via a path building up from a sequence of adjacent edges that (pair-wise) share at least one node~\cite{newman2018networks}. In temporal networks, however, connectedness is coded by {\em temporal} paths that are constructed from adjacent {\em temporal interactions}, which are not simultaneous yet structurally adjacent, and respect the causal time order. They determine the set of reachable nodes that can be influenced in the future with information held by a given node at a given time~\cite{Badie-Modiri2020,Holme2012}. The set of reachable nodes of a node at a given time, also called its influence set, is the node's temporal {\em out-component}, whose structure and size are important indicators of any ongoing dynamical processes. Indeed, no ongoing process can exhibit a larger collective pattern than the largest connected out-component in the underlying temporal network.
However, the characterisation of connected components in temporal networks is a difficult task, as the temporal ordering of interactions introduces a degree of complexity to detect time-respecting paths in an effective way.
Here, we address this challenge by defining a component matrix that codes the in- and out-component size of any node in a temporal network. Using this matrix we apply network compression and reconstruction techniques via graph hashing,
to estimate the distribution of the size of connected components of nodes. The proposed algorithm provides advancements in the computation efficiency of the largest node components compared to the state-of-the-art, specifically for temporal networks with large number of interactions.

\subsubsection*{Calculation of the largest out-component:}
Considering all nodes and timed interactions in a temporal network, the most important component to characterise is, among other components, the largest out-component that ever emerged in the structure. Its identification can be approached using different ideas.
A simple one would be to simulate a deterministic \emph{Susceptible-Infected (SI) process} starting from every node at their first interaction time. In a deterministic SI process, nodes are either in a susceptible (S) or infected (I) state and a susceptible node certainly becomes infected when interacting with an infected one. It is a conventional model to describe the fastest spreading process in a network, where starting from a single seed node at its first appearance, the downstream set of infected nodes determines its maximum out-component.
Using this method, in a temporal network of $n$ nodes and $m$ events, the computation of the out-component of a spreading seeded from a single source node, at its first appearance time would have $O(n)$ space and $O(m)$ time complexity (in terms of memory usage and computation time). This results in $O(n^2)$ space and $O(nm)$ time complexity when considering every node.

A more efficient method rely on temporal \emph{Event Graphs} (EG), a higher-order representation of temporal networks~\cite{Mellor2017,kivela2018mapping,Badie-Modiri2020}. An EG is a static and lossless representation of a temporal network in the form of a weighted and directed acyclic graph (DAG). In this structure, temporal interactions are associated to nodes that are linked if their corresponding events are adjacent.
For a more precise definition see Section.~\ref{sec:prob_def}.
Computing a single traversal of a static event graph (in reversed time order) yields the out-component of any node at any time, with an evidently smaller computational complexity as compared to a direct computation on a temporal network. However, EG appears with considerably larger size (having as many nodes as events in the original temporal network) and higher link density (by connecting any events to all future adjacent others) that leads to increased memory complexity. In order to reduce memory complexity, a link reduction method has been proposed that eliminates path redundancy in the EG~\cite{Mellor2017,kivela2018mapping}, leaving the connectedness of the DAG intact. Relying on the reduced EG, the use of the approximate HyperLogLog (HLL) counting algorithm can further reduce the time complexity of the out-component detection to $O(m \log(m)+\eta)$, where $\eta$ is the number of edges of the EG. However, this method provides only estimation for the size of out-components, without giving any information about their detailed structure.

\subsubsection*{Graph compression for component inference:} Contrary to earlier solutions, our idea is to use graph compression methods to compute the out-component size distribution of a temporal network, with a reduced computational complexity.
The compressibility of static networks has been studied recently \cite{Lynn2021}, and has been shown to depend on the structure of the graph. This notion can be extended for temporal networks by interpreting them as a sequence of time-aggregated static network snapshots. Then compression can be formulated as finding a smaller diffusion-equivalent representation \cite{Adhikari2017}. Also, consecutive snapshots can be compressed depending on their chronological importance~\cite{Allen2022}. Moreover, as pointed out by Li et al.~\cite{Li2017}, in spatio-temporal networks nodes can be compressed via local clustering, while reducing time instants to change-points. Compression can be formulated using Minimum Description Length to also reduce the size of the graph \cite{Liu2018}. Another compression approach has been proposed using information theory considerations, aiming to reduce the number of bytes required to describe a temporal network \cite{Panagiotis2022,Caro2016,Bernardo2013}. Reducing the size of the network via coarsening to compute spectral properties of a graph has also been studied \cite{Loukas2018}. Sampling techniques have been largely used to reduce the complexity of computation over large graphs \cite{Yousuf2020}.

Despite these numerous compression techniques proposed for temporal networks, none of them reduces effectively the number of nodes in a series of events. This reduction has a huge impact on the computational complexity of any of these algorithms, especially when they are characterised by quadratic complexity in the number of nodes. Thus our central question remains: \emph{how to design an efficient compression scheme that reduces the number of nodes while keeping enough information about the network itself to reconstruct the statistics of its connected components?} 

\vspace{.2in}

To reduce the computational complexity of the out-component size distribution calculation, we first propose a \emph{online streaming matrix algorithm} that scans through the series of events only once, while it can also consider new events added later on, without re-starting the computation. In addition, we define a general purpose temporal network compression scheme using a graph hashing approach. This compression method reduces the total number of nodes, yet it requires a decompression scheme too, which provides only an approximate solution. The compression method can be used in conjunction with the matrix algorithm and, more generally, it can be applied on any temporal network algorithm. 

To present our contributions, we organised the paper as follows. First, we formalise the problem of out-components computation in Section~\ref{sec:prob_def}. We present the proposed novel streaming matrix algorithm to compute the distribution of the size of out-components in Section~\ref{sec:our_method}, including some numerical experiments. Then, we describe the hashing framework in Section~\ref{sec:hashing}, and we report also on the numerical studies carried out to evaluate its ability to estimate the ground-truth out-components' distributions in Section~\ref{sec:experiments}. Finally, we discuss the proposed methods and the results.

\section{Methods}
\label{sec:prob_def}

The aim of the present work is to effectively compute the distribution of the maximum out-component size for all nodes in a temporal network. To establish our apporach, we introduce first the definitions that are necessary to ground our methodology.

\subsection{Problem definition}

We define a temporal network $\G:=(\V, \E, \T)$ as a series of temporal events $e = (u, v, t) \in \E$ that record interactions between nodes $u,v \in V$ at time steps $t$  sampled\footnote{In these definitions, we neglect the duration of events for simplicity, but all definitions could incorporate durations in a straightforward way.} from a $\T$ time periods of length  $T$. The network $\G$ is characterised by its number of nodes $n=|\V|$ and its number of events $m=|\E|$.
In $\G$ we call two events $e_i \in \E$, $e_j \in \E$ adjacent if they share at least one node $(\{u_i, v_i\}\cap \{u_j, v_j\} \neq \emptyset)$ and their inter-event time is $\Delta t=t_j-t_i>0$, i.e. the two events are not simultaneous. Furthermore, we call two events to be $\delta t$-adjacent if they are adjacent and their inter-event time is $\Delta t \leq \delta t$. A sequence of adjacent events defines a time respecting path between nodes $u$ and $v$ starting at time $t$, if the first event of the path starts from node $u$ at time $t$, the last ends at node $v$, and each consecutive events in the sequence are pairwise adjacent~\cite{Holme2012}. The set of nodes that can be reached by any paths starting from node $u$ at time $t$ defines the out-component.
The size of the out-component of a node $u$ at a given time $t$ is measured as the number of unique nodes that can be reached by valid time respecting paths. Actually, it determines the largest possible phenonenon (e.g., largest epidemic or information cascade) that was initiated from that source node and evolved in the future. The computation of out-components is computationally challenging as it requires the tracking of each time-respecting paths starting from each node at each time. However, an effective approximate solution has been proposed lately~\cite{Badie-Modiri2020} to solve a partial challenge, to estimate only the size of out-components without keeping track of nodes involved.

\subsection{Event graphs and the HyperLogLog algorithm}


The proposed solution builds on the Event Graph (EG) representation~\cite{kivela2018mapping,Mellor2017} of temporal networks.
An event graph $G:=(V, E, \Delta t)$ is defined as a static weighted directed acyclic graph (DAG) representation of a temporal network $\G$, where temporal events are associated to nodes in $G$ (i.e., $V=\E$); directed edges in $G$ correspond to $\Delta t$-adjacent event pairs in the original temporal network, with direction indicating their temporal order. The $\Delta t$ weight of each link is defined as the inter-event time between the two adjacent events corresponding to the connected nodes in $G$.
This way, an event graph has $m=|\E|$ number of vertices and $\eta$ number of directed edges. This static graph representation provides a losless description of a temporal network and can be exploited to infer several property of $\G$ without computations on the temporal structure~\cite{kivela2018mapping}. Indeed, thanks to the EG representation, the out-component size distribution of $\G$ can be precisely computed~\cite{Badie-Modiri2020}, yet with high computational and memory costs.

To reduce this cost at the price of an inexact computation, Modiri et al.~\cite{Badie-Modiri2020} proposed an approximate solution to precisely estimate the out-component size distribution of a temporal networks using its EG representation combined with the HyperLogLog algorithm.

The HyperLogLog (HLL) algorithm takes as input a set, and it outputs an approximate of its size \cite{Flajolet2007}. More precisely, a HLL structure uses a representation on $s$ registers each storing a number, initialised to zero to start with. Every element of the set is hashed into a binary vector that is then cut in two parts. The first part indicates the identifier of the register that will be used and the position of the leftmost $1$ in the second part is stored in that register if it is larger than the current value. Finally, the size of the set is estimated with an ensemble indicator function based on the registers. The main advantage of the algorithm is that the whole set is not stored to estimate its size and the estimation can be done with constant space and time complexity $O(s)$. The error of the estimation is $O(1/\sqrt{s})$. Also, the size of the union of two sets can also be estimated in constant time and space by merging two HLL structures. A final property is that each element of the set is considered one by one, hence compatible with a streaming approach. Let us stress that the hashing functions used in HLL are not related to the ones we will use in Section \ref{sec:hashing} to compress the network representation.

The HLL algorithm can be used to estimate out-component sizes in a EG without tracking the exact set of nodes involved~\cite{Badie-Modiri2020}. This approach reduces the time complexity of the out-component distribution computation to $O(m \log(m) + \eta)$ up to some constant factors that depend on the hyper parameters $s$ of the HLL algorithm, which sets the trade-off between computational efficiency and accuracy.

\section{Streaming matrix algorithm for out-component size calculations}
\label{sec:our_method}

We develop a streaming matrix algorithm as an exact solution for the question of computing the largest out-component of each node in a temporal network. The proposed solution can process chronologically streamed nodes and events of a temporal network in real time, with a space complexity that does not depend on the number $m$ of events.

To demonstrate the basic idea of the method, let us consider the simple example of an information spreading process on a temporal network between $n$ nodes modelled by a deterministic SI process (a short definition is recalled in the Introduction). To follow-up on the evolving components during the SI process, we design a matrix with rows representing the in-component and columns representing the out-component of each node. At time $t=0$, when each node has a unique information that it has not propagated yet to any other nodes, we obtain the identity matrix with ones in the diagonal and zeros otherwise. Propagation happens between nodes $u$ and $v$ at the time of their interactions, when they mutually share all unique information they already learned from others (including their own) during earlier times of the process\footnote{Information sharing could be deemed non-mutual in case of directed interactions in the temporal network.}.
This propagation rule is associated to the "OR" operation between the corresponding lines of the matrix, which yields the {\em union} of the set of unique information known by the two nodes. 
By the last event of the temporal network, the unique information of a node $u$ is known by all other nodes in its out-component (depicted by column of the matrix). Thus, to compute the size of $u$'s out-component, we simply have to count the number of unique nodes that are aware of $u$'s unique information, i.e., the number of ones in the corresponding column of the matrix. 

\subsection{The component matrix}

\begin{figure}[!h]
    \centering
    \includegraphics[width=\textwidth]{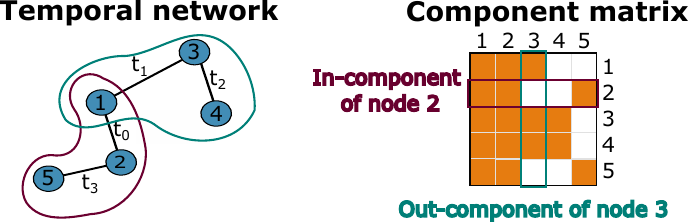}
    \caption{Left: a temporal network defined as a series of events ordered in time ($t_i<t_j$ for $i<j$). The out-component of node 3 is circled in blue; the in-component of node 2 is circled in purple. Right: the corresponding  component matrix. A row depicts the in-component of a node (we emphasise that of node 2). A column depicts the out-component of a node  (we emphasise that of node 3): a non-zero element in the $u$-th column at coordinate $v$, means that node $v$ belongs to the out-component of node $u$. }
    \label{fig:matrix_algo_introduction}
\end{figure}

The component matrix is a binary matrix of size $n \times n$, where $n$ is the number of nodes in $\G$. An illustration is provided in Fig.~\ref{fig:matrix_algo_introduction}. An element $(i, j)$ of the matrix is $1$ if and only if the node $i$ is reachable from the node $j$ by any temporal path. Thus, the $i$-th line of the component matrix is the in-component of the node $i$ and the $j$-th column of the matrix is the out-component of the node $j$.  

\begin{algorithm}
\SetAlgoLined
\caption{The matrix algorithm}\label{alg:matrix_algo}
\KwData{Time ordered list of events $E$}
\KwResult{Component matrix $S$}
$S \gets I_n$ \tcp*{Initialization}

\For{$e$ in $\E$}
{ 
  u $\gets$ e[0] \\ 
  v $\gets$ e[1] \\ 
  r $\gets$ S[u] OR S[v] \tcp*{Compute the binary OR}
  S[u] $\gets$ r \tcp*{Replace row of $u$ by $r$}
  S[v] $\gets$ r \tcp*{Replace row of $v$ by $r$}
  }
\end{algorithm}

\begin{figure}[!h]
    \centering
    \includegraphics[width=\textwidth]{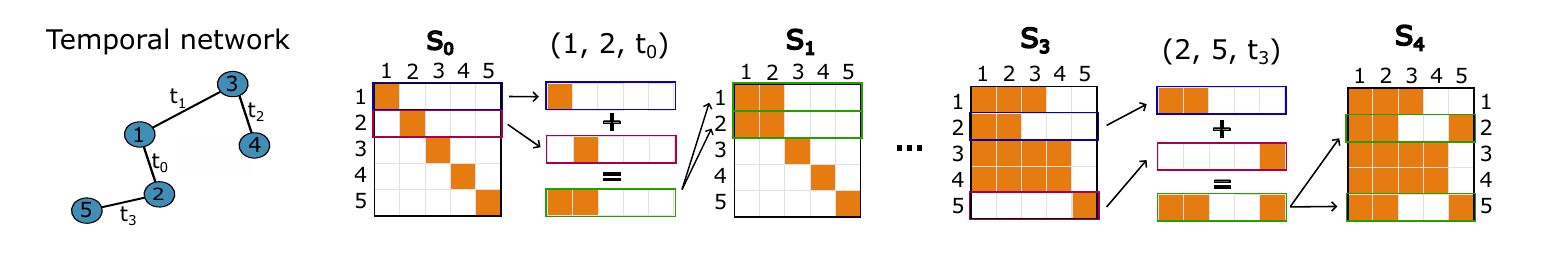}
    \caption{From a given temporal network (left graph), we compute the Component Matrix (right matrix) of size $n \times n$, with $n=5$, by scanning the series of events. For each event, we compute the OR operation between the rows of the matrix corresponding to the interacting nodes and replace them by the result. Matrix $\mathbf{S_4}$ is the component matrix at the end of the streaming after $m=4$ events.  }
    \label{fig:matrix_algo}
\end{figure}

The precise algorithm to compute this component matrix is given as pseudo-code in Algorithm~\ref{alg:matrix_algo}. It starts with the identity matrix. Then, for every event, the lines corresponding to the interacting nodes are used to compute a binary OR operation, and those lines are replaced by that resulting OR. Finally, at the end of the series of events, the output matrix is the component matrix. This algorithmic construction process is described in Fig.~\ref{fig:matrix_algo}.

\subsection{Complexity of the algorithm}

Since we use a $n \times n$ matrix to store the intermediate results, the space complexity is $O(n^2)$, which may be reduced using sparse matrices for storage. For time complexity, we can divide the algorithm into several steps. The initialisation of the identity matrix can be done in $O(n)$ by simply setting the $n$ diagonal elements to the "True" value at the outset. To update the matrix we perform the OR operation between two vectors of size $n$ once for each of the $m$ events. The complexity of each update is $O(n)$, bounded by the maximum out-component sizes $n$, but could be further reduced with a sparse matrix format. Consequently, the total complexity of the updates is $O(nm)$. Finally, counting the number of non-zero element (or non-"False" elements) can be done at the same time as the update without any added complexity. Thus, the overall time complexity of the component matrix algorithm is $O(n) + O(nm) = O(nm)$.

\subsection{Streaming computation of average out-component size}

One way to further reduce the complexity of the computation is to look for \emph{an approximation} of the \emph{average size} of a maximum out-component rather than its exact size. This can be implemented with the component matrix using the HyperLogLog counting algorithm that has been recalled before. It allows us to approximately describe and count the "True" values on each row of the matrix. The rows of the component matrix describe a set of nodes: the in-component. An HLL structure of size $s$ (arbitrarily chosen, independently of $n$ and $m$) can be used to estimate the size of an in-component. Thus, $n$ HLL structures replace the former matrix. In its matrix form, the algorithm starts with the identity matrix. For the HLL structures, we simply initialise them with a single element: the $i$-th structure will be initialised with "$i$". Then, for every event $(i, j, t)$, the OR operation between the lines $i$ and $j$ of the matrix is computed, which is equivalent to the union of the two in-components. For the HLL structures, this results in merging them. Finally, every HLL structure can give an approximation of the size of its corresponding line in the component matrix.

Interestingly, the average size of the maximum out-components at time $t$ is defined as:
\begin{equation}
    \Bar{s}_t = \frac{1}{n} \sum_{u \in \V} \sum_{v \in \V} S_t(u, v),
    \label{eq:s_t}
\end{equation}
where the sums are interchangeable. Thus it can be computed both as the average size of out-components or {\em in-components}. Actually, the HyperLogLog structure can compute a size estimate with no additional cost in $O(1)$ time complexity for each in-component, which are coded in the matrix as the number of "Trues" in a row. According to Eq.~\ref{eq:s_t}, the average value of these maximum in-component size values can give us an estimate directly for the average of the maximum out-component sizes. Thus, the HyperLogLog approach can reduce the algorithm's space complexity to $O(n)$ and the time complexity to $O(m)$~\footnote{Remember, however, that the $O(\cdot)$ notation hides a constant, whose value results from a trade-off between cost and precision for the HLL algorithm.}. As an advantage, the matrix algorithm using HyperLogLog preserves the streaming aspect of the algorithm, assuming events to arrive in chronological order. However, in turn, it does not provide the whole maximum out-component size distribution but only an estimation for its mean value.

\subsection{Component size distribution from reversed event sequence}
\label{sec:3_4}

By reversing time, we can easily obtain a solution to compute the whole maximum out-component size distribution. But it comes at the expense of losing the streaming property of our algorithm, as this solution takes as input, the whole interaction sequence in reversed order, processing it from the end to the beginning. By reversing the order of the sequence of events, the in-components become the out-component. 
In this case the component matrix algorithm does not fuse the rows anymore but it has to be adjusted the fuse the columns instead.

Thanks to the reversal of the sequence of events, we can use the HyperLogLog counting method to estimate the full distribution of the maximum out-component sizes at a lower cost.

More specifically, for every node, we initialise a HyperLogLog structure with constant size, which contains only the node itself as previously. Then, for every event $(u_i, v_i, t_i)$, considered {\em in reverse chronological order}, we merge the structures of $u_i$ and $v_i$ (corresponding to columns, i.e., to current estimates of out-components) in $O(1)$ time. Finally, we approximate the size of the maximum out-component of every node with their HyperLogLog estimates. This results in having an approximation for the whole distribution of out-components' sizes in $O(n)$ space complexity, $O(m)$ time complexity, and scanning the events' sequence only once. While this seems to be a very efficient solution, the constants in the complexity evaluations are quite large in practice, setting back the effective performance of this solution in some regimes of $n$ and $m$, as we demonstrate in the next section, while providing better results in others. 

\begin{table}[!t]
    \centering
    \begin{tabular}{|c|c|c|c|c|c|c|}
    \hline
        Method & Time cpx. & Space cpx. & Exact & Stream & OC & P(|OC|).\\
        \hline
        EG + HLL & $O(m\log(m) + \eta)$ & $O(m+\eta)$ & No & No & No & Whole\\
        SI process & $O(mn)$ & $O(n^2)$ & Yes & No & Yes & Whole\\
        \hline
        Matrix & $O(mn)$ & $O(n^2)$
        & Yes & Yes & Yes & Whole\\
        Matrix + HLL & $O(m)$ & $O(n)$ & No & Yes & No & Average \\
        Matrix + reverse t & $O(mn)$ & $O(n^2)$ & Yes & No & Yes & Whole \\
        Matrix + reverse t + HLL & $O(m)$ & $O(n)$ & No & No & No & Whole \\
        \hline
        \hline
        Matrix + hashing (//) & $O(mn_s)$ & $O(n_s^2K)$ & No & Yes & Yes & Whole \\
        Matrix + hashing & $O(mn_sK)$ & $O(n_s^2)$ & No & Yes & Yes & Whole
        \\
        \hline
    \end{tabular}
    \caption{Summary of methods used to compute the maximum out-component size distribution. Time and space complexity depends on the number of nodes $n$ and events $m$, and number of edges $\eta$ in the event graph (EG). Column entitled "Exact" indicates if the method provides exact (Yes) or approximate (No) solution. The column called "Stream" indicates if the method can stream events in chronological order. Column "OC" shows if the method can compute not only the out-component sizes but the involved nodes as well. The column "$P(|OC|)$" shows if the whole out-component distribution (Whole) or only its average (Average) can be computed. The hashing framework is described in the next section, with $n_s$ number of super-nodes, $K$ number of hashing functions, and // indicating the possible parallelisable method. 
    Note about the Matrix method: its space complexity is in $O(n^2)$ but 
    can be reduced to $O(\bar{s} n \log n)$ where $\bar{s}$ is the average size of out-components in the case of sparse matrices.
 }
    \label{tab:complexities}
\end{table}

As a summary, the first part of \Cref{tab:complexities} reports on the complexity and properties of the methods described so far. The second part of the table also anticipates on the method to be described in the next section.

\subsection{Experimental validation}

We perform several computational experiments to demonstrate the effectiveness of the component matrix algorithm and to compare its performance to the corresponding EG based solution.

\subsubsection{Experimental setting: }

To test the performance of these algorithms we consider a simple model of temporal network. We first generate a static undirected random graph $G(n,p)$ using the Erd\H{o}s-Rényi model with $n$ nodes and wiring probability $p=2/n$. This way the constructed static graph will likely contain a unique giant component. To generate a temporal network, we set an independent Poisson process on each link~\cite{Badie-Modiri2020} of this graph to determine the times when links are present and can transmit information between connected nodes. This way, both the underlying structure and the link dynamics are generated by random processes, that induce limited degree heterogeneity (with an emerging Poisson degree distribution)~\cite{albert2002statistical} and no burstiness (with exponential inter-event time distribution)~\cite{karsai2018bursty}. In the simulated networks, the number of nodes $n$ varies between $\{100, 200, 500, 1000, 2000, 5000, 10000\}$ and the number of events $m$ logarithmically from $10$ to $10^8$. Note that while real temporal networks may exhibit several types of structural and temporal heterogeneity, we assume they would not change considerably the conclusion of the performance evaluation provided here.

For a fair comparison, both the EG based and the component matrix based methods were used to solve the same task, that is to compute the largest maximum out-component size of a temporal network. While the EG based method solves a larger problem first, i.e. estimating the out-component size of any node at any event, one can extract the maximum out-component size for every node from its solution, simply by taking the size that corresponds to the first emergence of a given node. The overall asymptotic memory and time complexity of this solution scales similarly to the EG+HLL algorithm~\cite{Badie-Modiri2020}, as it is summarised in Table \ref{tab:complexities}. Taking this model as reference, we compare it with the performance of the proposed proposed method based on the exact component matrix, as well as with its variant which uses HLL algorithms to obtain an approximation. In each method using HLL, we tune $s$ 
to obtain less than $1\%$ error for the average component size. Note that reversing time does not change neither the time nor the memory complexity of the component matrix algorithm with or without HLL as summarised in Table \ref{tab:complexities}.

\begin{figure}[!h]
    \centering
   	\includegraphics[width=.85\textwidth]{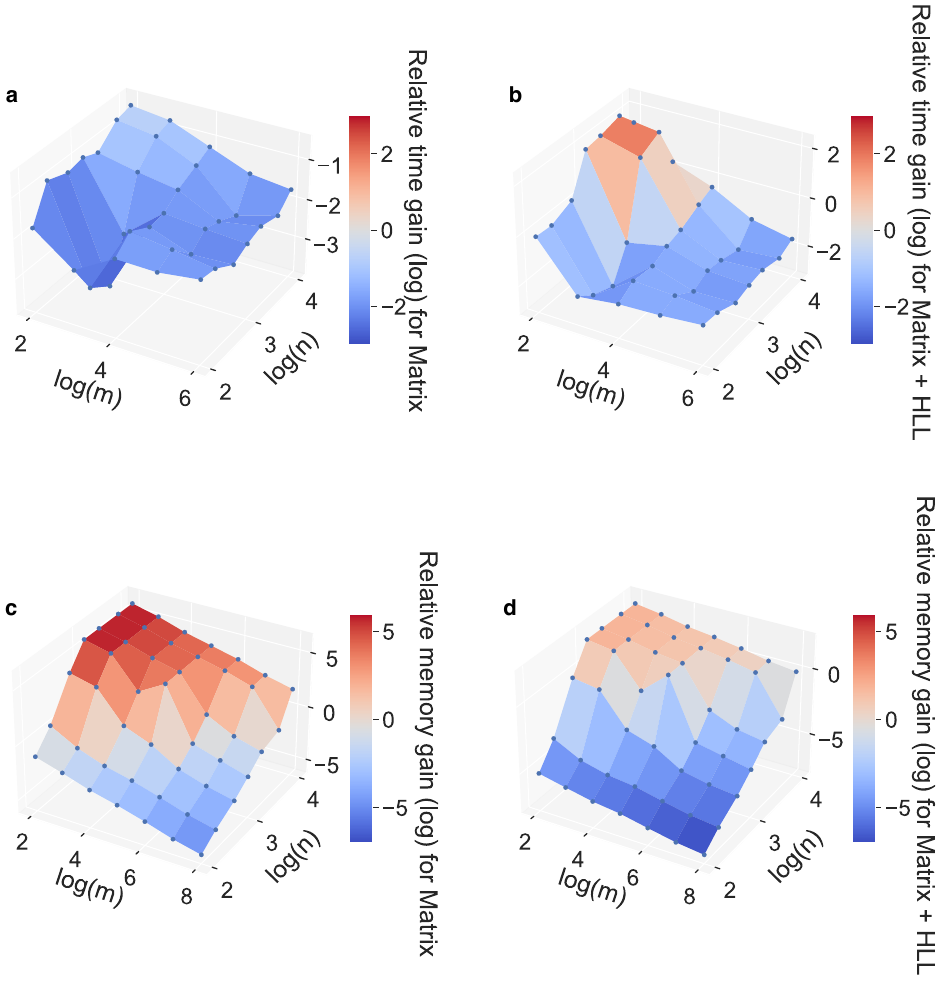}
    \caption{Fraction of computational time (top row) and memory usage (bottom row) of the component matrix methods divided by the ones of the EG+HLL method. Panels (a) and (c) depict results for the exact component matrix method, while panels (b) and (d) are for its approximate solution using HLL. All scales are logarithmic with colour blue indicating when the component matrix method performs better (and red in the contrary case). 
    }
    \label{fig:comparison_with_eventgraph}
\end{figure}

\subsubsection{Results: }

To compare the different methods, we report in Fig.~\ref{fig:comparison_with_eventgraph} the ratio of computation times  and of memory usages between the compared algorithms. First, let us focus on the relative performance of the component matrix method in Fig.~\ref{fig:comparison_with_eventgraph} (a) and (c). Interestingly, results depicted in panel (a) suggest that although this method provides an exact solution for the task, it performs always better than the EG+HLL algorithm in terms of computation time. A similar scaling is true in terms of memory complexity (panel (c)), although the large quadratic cost of the component matrix makes this method to perform worst than the reference for small numbers of events or for large numbers of nodes. Nevertheless, we can conclude that the component matrix method largely outperforms the event graph method in terms of computational time and memory for large numbers of events, especially with networks of smaller size, where the gain can reach several orders of magnitude.

Comparing the HLL variant of the component matrix algorithm with the reference EG+HLL mode, show more variable performance. In terms of computational time (see panel (b) in Fig.~\ref{fig:comparison_with_eventgraph}), although worse for small numbers of events and large networks, the performance of our method is comparable to that of the exact method, for the other parameter values. However, regarding memory consumption (see panel (d)), our method is much more efficient, as it does not have to store the component matrix of size $n^2$. For large network sizes the model requires approximately the same memory size as the reference model, while it is doing significantly better for the rest of the parameter space.

\subsubsection{Advantages and limitations:}

As stated before, a major advantage of the component matrix algorithm as compared to other methods is its space complexity that does not depend on the number $m$ of events in the temporal network but scales as the square of its node set size $n$. Meanwhile, its time complexity scales only linearly with $m$. This is especially suitable for data streaming scenarios when nodes and events arrive in chronological order.
Actually, adding a new node to the network requires only to add a new row and column to the component matrix set as "False", except the diagonal element. As for new events, insertion follows the update rule discussed earlier, as the algorithm operates in a streaming manner. Furthermore, the component matrix method requires only one pass over the event sequence. At any time step $t$ when a new event appears, it only requires information about the previous state of the component matrix $S_{i-1}$ at time $t-1$ (or conversely  in the case of reversed time).
On the other hand, the exact component matrix method scales poorly in space complexity in terms of $n$, the number of nodes, as it operates on a $n \times n$ matrix. This shortcut can be addressed by the HLL method to obtain approximate results. A sparse matrix implementation 
can also be very beneficial to solve this problem, if the average out-component size is much smaller than $n$. Otherwise, when it is comparable to $n$ and the number of non-zero elements in the matrix is in $O(n^2)$, even a sparse matrix solution would scale quadratically. 

\subsubsection{Reference point:}

The figure above only gives the ratio between the computation times or the amount of memory required for the computations. We provide in Table~\ref{tab:ref_point1} some reference points for the real values for each method. The smallest temporal network on Fig.~\ref{fig:comparison_with_eventgraph} is for $n=100$ and $m=100$. 
The largest temporal network in Fig.~\ref{fig:comparison_with_eventgraph} is for $n=10^4$ and $m=10^8$. The associated computation times and memory usages are reported in Table~\ref{tab:ref_point1}.

\begin{table}[t]
    \centering
    \begin{tabular}{c|c|c|c}
         & EG + HLL & Matrix & Matrix + HLL \\
        \hline
        Time (ms) & 40.3 & 0.4 & 27.4\\
        Memory (kB) & 1 & 80 & 0.92 \\ 
        \hline
        Time (s) & 30172 & 700 & 811\\
        Memory (MB) & 200000 & 800 & 0.085 
    \end{tabular}
    \caption{Computation times and memory usage for the EG + HLL, Matrix and Matrix + HLL methods for $n=100$ and $m=100$ (first and second line) and for for $n=10^4$ and $m=10^8$ (third and fourth line). }
    \label{tab:ref_point1}
\end{table}

\section{Hashing the temporal network}
\label{sec:hashing}

\begin{figure}[!h]
    \centering
    \includegraphics[width=\textwidth]{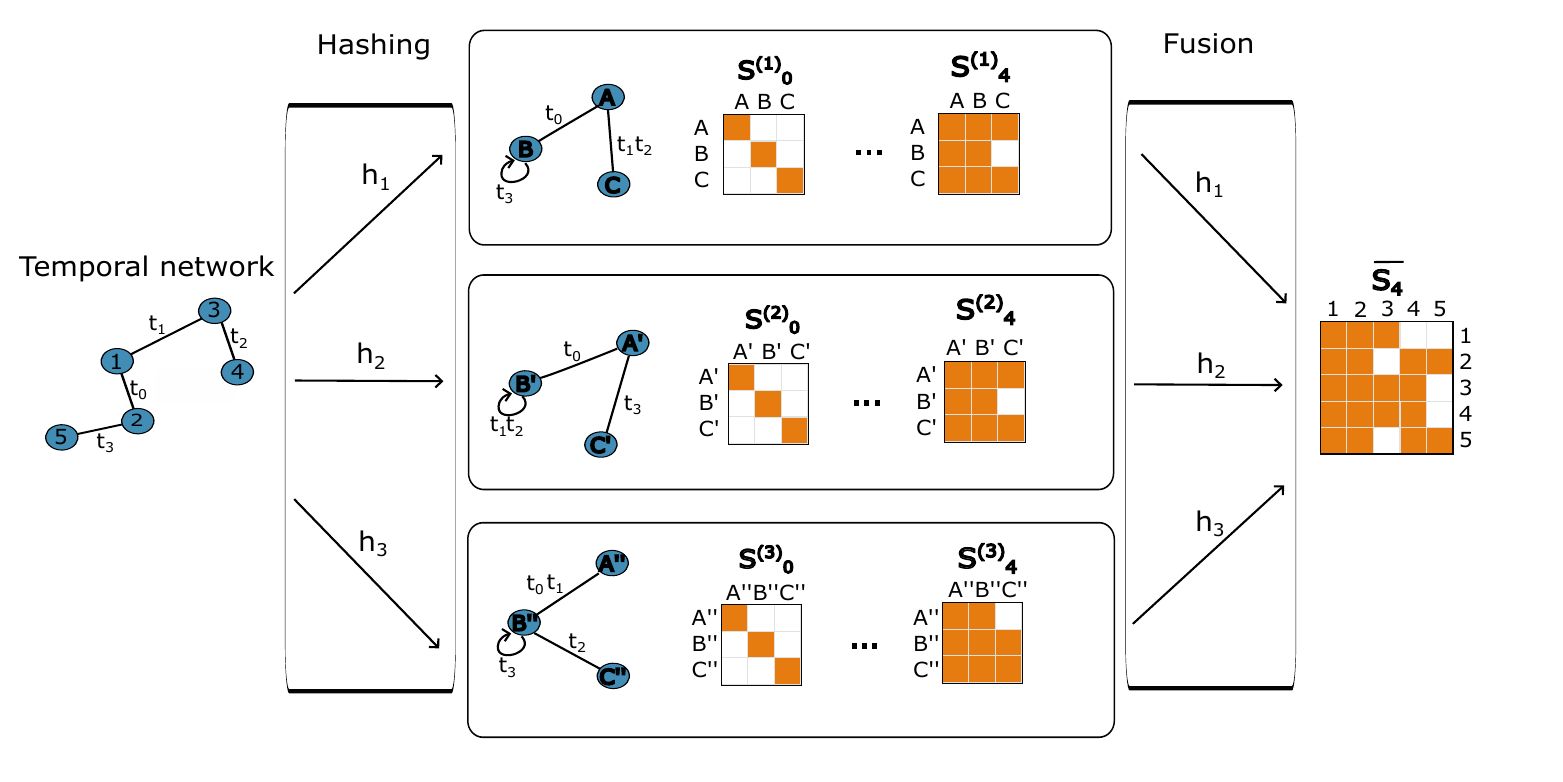}
    \caption{From a $5$ nodes temporal network, several hashed version are computed with $3$ nodes each. Then, every hashed graph can be used to compute a small component matrix thanks to our matrix algorithm. Finally, the different component matrices can be fused to compute an approximate solution of the component matrix of the initial temporal network. 
    }
    \label{fig:hashing}
\end{figure}

Hashing the temporal network consists in reducing its number of nodes, thus compressing it, by (randomly) assigning nodes of the initial temporal network to "super-nodes" of a hashed graph. An event, or an interaction, between two nodes at time $t$ in the initial temporal network becomes a new event between their hashed representatives in the hashed graph at time $t$. 
Reducing the number of nodes is notably attractive 
because this reduces the complexity of various different algorithms including the computation of the component matrix, even though 
this may cause  information loss about the initial graph. To balance this effect, we propose to use several different hashing functions and to fuse the obtained results together. The overall framework is shown in Fig.~\ref{fig:hashing}.

\subsection{Hashing functions}

To reduce the number of nodes of the static graph underlying the input temporal graph (in short: "the input static graph"), and therefore the computation complexity of  the out-component size distribution,  we use hashing functions. These functions take as input a set of labels of $n$ nodes, $\{1, ..., n\}=[n]$, and hash them into $n_s$ super-nodes $\{1, ..., n_s \}=[n_s]$. The labels are the nodes of  the input static graph and the buckets are the super-nodes of the resulting hashed static graph. Since $n_s<n$, some nodes will collide into the same super-node, reducing the overall cost of computation over the hashed temporal network associated to the hashed static graph, but reducing also the amount of available information. 

We use $k$-universal (randomised) hashing functions \cite{Thorup2004}. A \PG{}{with $k=4$. In general, a} class $H$ of random functions is $k$-universal if $\forall x_1, ..., x_{k} \in [n], \forall v_1, ..., v_{k} \in [n_s],$ 
\begin{equation}
    \text{Pr}
    \{ h(x_i)=v_i, \forall i \in [k]\} = 1/n_s^k
\end{equation}
where the probability is on the draw of $h$. 
Qualitatively, this means that, the probability that one node of the initial graph is assigned to the same super-node by two different hashing functions is low and controlled by the choice of $k$.

In our work, we use $k=4$ and the hashing functions are based on a large prime number: $Prime=2^{61}-1$. First, let us define the table $A$ of size $(3, order)$, where $order$ is an order parameter, as:
\begin{equation}
    \forall i \in \{0, 1, 2\}, \forall j \in [order], A(i, j) = ((rand(2^{31}-1) \ll 32) + rand(2^{31}-1)) \% Prime
\end{equation}
where $\ll32$ denotes the shift of the binary representation $32$ bits to the left.

The number $acc$ is computed recursively $order-1$ times thanks to:

\begin{equation}
    acc(i, u) = MultAddMod(u, acc(i, u), A(i, j)) \% Prime
\end{equation}
where $j \in \{1, ..., order\}$, $MultAddMod$ is a function defined in the paper and initially, $acc(i, u) = A(i, 0)$.

Then, 
$T_0, T_1, T_2$ are tables of size $n_s$ that are defined as: 
\begin{equation}
    \forall u \in [n_s], T_i[u] = acc(i, u)
\end{equation}
Furthermore, for every node, we define three quantities $x_0, x_1, x_2$ as $\forall u \in [n], x_0(u) = low(u), x_1(u) = high(u), x_2(u) = x_0(u) + x_1(u)$ where $low(u)$ outputs the $32$ rightmost bits of the binary representation of $u$ and $high(u)$ outputs the $32$ leftmost bits.
Finally, 
\begin{equation}
    \forall u \in [n], h(u) = T_0(x_0(u)) \star T_1(x_1(u)) \star T_2(x_2(u))
\end{equation}
where $\star$ is the bitwise exclusive OR.

The hashed static graph is made of  super-nodes defined by the output of a hash function and of "super"-edges connecting them: if $u$ and $v$ are connected in the initial static graph, then the super-nodes $h(u)$ and $h(v)$ become connected by a super-edge, whose weight is binary. Finally, for every event $(u, v, t)$, a super-event is defined as $(h(u), h(v), t)$.

\subsection{Fusion to compute the distribution of out-components}

The main goal of our work is to compute the out-component, or its size, of every node in the input temporal network with lower complexity than existing methods reported in Table \ref{tab:complexities}. To do so, we hash the set of $n$ nodes of the input temporal network into $n_s$ super-nodes with $K$ 
different hash functions $h_j$. 
\begin{equation}
    \forall i \in [n]\ \forall j \in [K], h_j(u_i) \in [n_s]
\end{equation}
These hashing functions are drawn independently at random.

Here, the hashing functions ${h_j, \forall j \in [K]}$ are not injective thus not invertible: there are usually several nodes mapped to the same super-node. We define the inverse of $h$ as the function that, given a super-node of the hashed static graph, computes the set of corresponding nodes in the initial static graph: 
\begin{equation}
    \forall v \in [n_s],\ h^{-1}(v) = \{u \in [n]/ h(u)=v\}
\end{equation}

Denote $\mathtt{OC}(u)$ (resp. $\mathtt{OC}(h(u))$) the out-component of node $u$ (resp. super-node $v = h(u)$). Assuming we can compute (an estimate of) the out component $\mathtt{OC}(v)$ of a super-node $v$ in the hashed graph obtained with hashing function $h_j$, we can also define
\begin{equation}\label{eq:DefInvHComp}
    h_j^{-1}(\mathtt{OC}(v)) = {\bigcup_{x \in \mathtt{OC}(v)} h^{-1}(x)}
\end{equation}

Instead of estimating the out-component for each of the $n$ nodes in the temporal network, we first hash the network into $K$ hashed graphs of $n_s$ nodes and $m$ events, then estimate the out-component for every node in the hashed graphs and finally aggregate the information by intersecting the (estimated) out-components given by each hashed graph.
We then define 
\begin{equation}\label{eq:DefEstOCMerge}
    \forall i \in [n], \widehat{\mathtt{OC}}(u_i) = \bigcap_{j \in [K]} h_j^{-1}(\mathtt{OC}(h_j(u_i))).
\end{equation}

The estimated out-component necessarily contains the true out-component, i.e. $\mathtt{OC}(u_i) \subseteq \widehat{\mathtt{OC}}(u_i)$, yet if $K$ is too small the set $\widehat{\mathtt{OC}}(u_i)$ may be much larger than the true out-component. 
Computing $|\widehat{\mathtt{OC}}(u_i)|$,  where $|A|$ is the number of elements of a set $A$, one can compute an approximation of the distribution of the out-components' sizes with any of the aforementioned algorithm that is able to compute the out-components ({\em and not only their sizes}) on the hashed graphs. We compare the resulting approximate distribution with the true distribution.

\subsection{Properties of the algorithm}

The structure of the resulting algorithm ensures that every step before the final fusion remains compatible with streamed events arriving in chronological order, and is also amenable to parallel/independent computations for each hashing function. 

Moreover, the complexity of the framework depends on the setup. In a parallel setting, \textit{i.e.} when the $S^{(k)}_i$ are computed separately, we need $O(K \times n_s^2)$ space to store the matrices and $O(mn_s)$ time to compute the small component matrices. In a non-parallel setting, we need $O(K\times n_s^2)$ to store the small matrices and $O(Kmn_s)$ time to compute them. 

\subsection{Experimental evaluation}
\label{sec:experiments}

The compression framework that we propose can be used with several observables. Here, we focused on the computation of the out-components. The whole distribution of the size of the out-components describes the largest spreading phenomena possible starting from every node. The other quantity we are interested in is the tail of the distribution, \textit{i.e.} the set of nodes with largest out-component' size. To experimentally prove the effectiveness of our work, we measure the precision of the approximate method with respect to the ground-truth for both the distribution and its tail. 

\subsubsection{Experimental setting: }

For simulations, temporal networks are generated exactly as in the previous Section, see \ref{sec:3_4}.
In the generated data, the number of nodes, $n$, varies between $\{100, 200, 500, 1000, 2000, 5000, 10000\}$ and the number of events varies between $10^4$ to $10^9$ as powers of $10$. The number of super-nodes is always a fraction of the number of nodes: $n_s = 0.3 \times n$ and the number of hashing functions is $K=5$.

The baseline algorithm is our matrix method from the previous section since it provides the exact distribution of the size of the out-components, with a controlled memory and time complexity. The results will compare the hashing version to this baseline.

In addition, some experiments have been conducted on real-world datasets freely available in \href{http://snap.stanford.edu/data/}{Snap}. The "Superuser" temporal network is a network of interactions on the stack exchange web site Super User. There are three kinds of interactions (edges): a user (node) answered a question of someone else, a user commented a question or a user commented an answer. 
The Superuser network is made of $194\,085$ nodes and $1\,443\,339$ events. The "Reddit" dataset is a temporal network of connections between subreddits (i.e., forum dedicated to a specific topic on the website Reddit), taken here as nodes.
There is an event between two subreddits when there is a post from a source subreddit that links to a target subreddit. The Reddit temporal network has $35\,776$ nodes and $286\,560$ events. 
 
For real datasets, we split chronologically the events in $10$ equal parts and compute the distribution of the largest out-component on $\{10\%, 20\%, ..., 100\%\}$ of the events. As for generated datasets, we use $n_s = 0.3 \times n$. We also use $K=1$ and $K=5$.
The baseline is still the matrix algorithm of the previous section.

\subsubsection{Performance criteria: }

We evaluate the hashing framework based on three criteria: time, memory and accuracy. We compare the time required by the matrix algorithm to compute the true distribution of the largest out-components, $\mathcal{D}$ with the one required by the hashing framework to compute the approximate distribution of the largest out-components $\mathcal{D}_s$.
The computation time of the hashing framework includes both the computation of the hashed matrices and their fusion. 

We also compare the memory usage of the matrix algorithm, with a single big matrix, with the one of the hashing framework, with several smaller matrices. 

Furthermore, to compare the ground-truth out-component' size distribution $\mathcal{D}$ and the one computed thanks to the hashing framework $\mathcal{D}_s(n_s, K)$, we simply use the Earth-Mover distance, also called Wasserstein distance \cite{earth_mover}, computed thanks to the Python Optimal Transport library \cite{flamary2021pot}, but any distance between two distributions could be used. We also tried to use the Kullback-Leibler divergence but it was less sensitive to subtle differences between the distributions. Thus we can define accuracy of the out-component size distribution inference as:
\begin{equation}
    Acc(n_s, K) = \gamma (\mathcal{D}, \mathcal{D}_s(n_s, K))
\end{equation}
where $\gamma$ 
is the Earth-Mover distance. The lower $Acc(n_s, K)$, the closest are the two distributions.

\begin{figure}[ht!]
    \centering
    \includegraphics[width=0.95\textwidth]{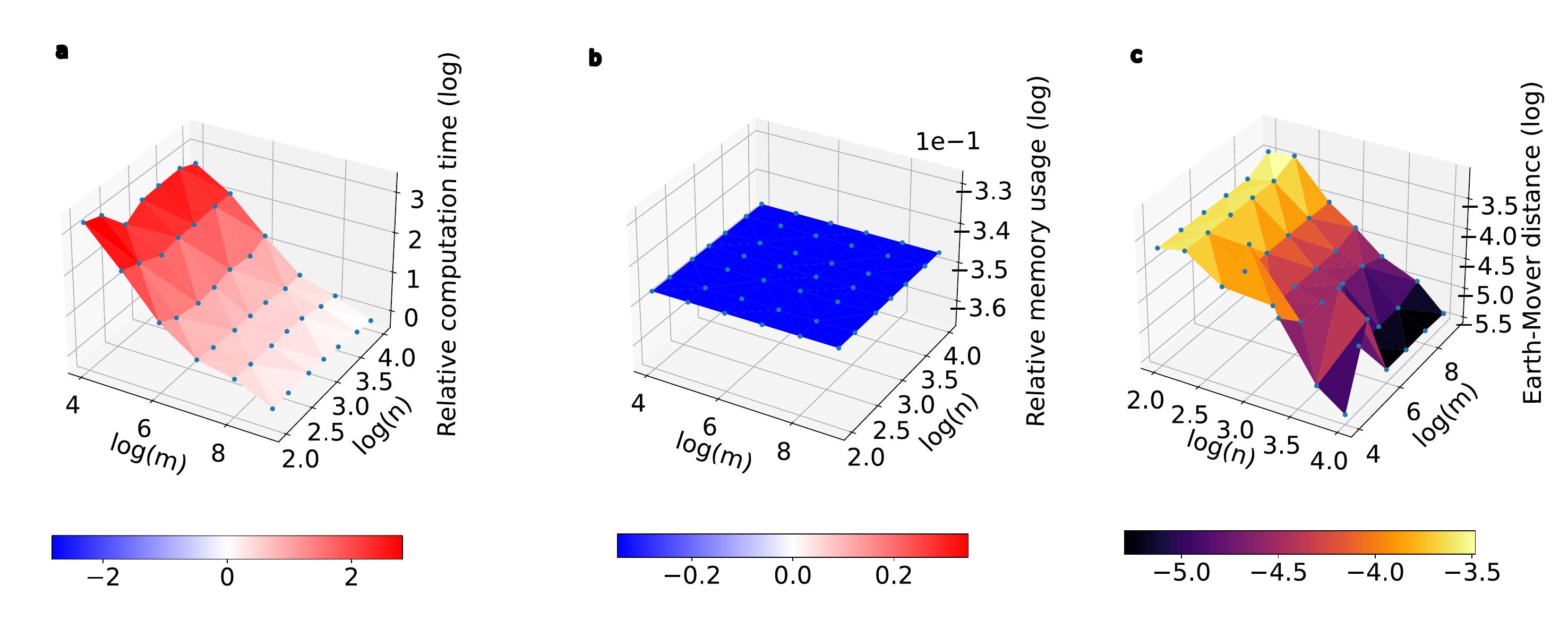}
    \caption{a) Relative time and b) relative memory usage for the computation of the distribution of the largest out-components compared to the ground-truth given by the matrix method on synthetic data. c) Accuracy of the hashing framework given by the Earth-Mover distance.}
    \label{fig:hashing_generated_data}
\end{figure}

\subsubsection{Results}

First, we present the result for the generated data.  The relative computation time, relative memory usage and accuracy of the hashing framework are reported in Fig.~\ref{fig:hashing_generated_data}.
The relative computation time figure is red meaning that the hashing framework requires more time than the matrix method to compute the target distribution. However, we can clearly see that the relative computation time decreases quickly with the number of events and slowly with the number of nodes. Generally, for datasets with more events than $m=10^8$, the hashing framework with $K=5$ and $n_s=0.3 \times n$ requires less time than the full matrix method.

\begin{figure}
    \centering
    \includegraphics[width=.65\textwidth]{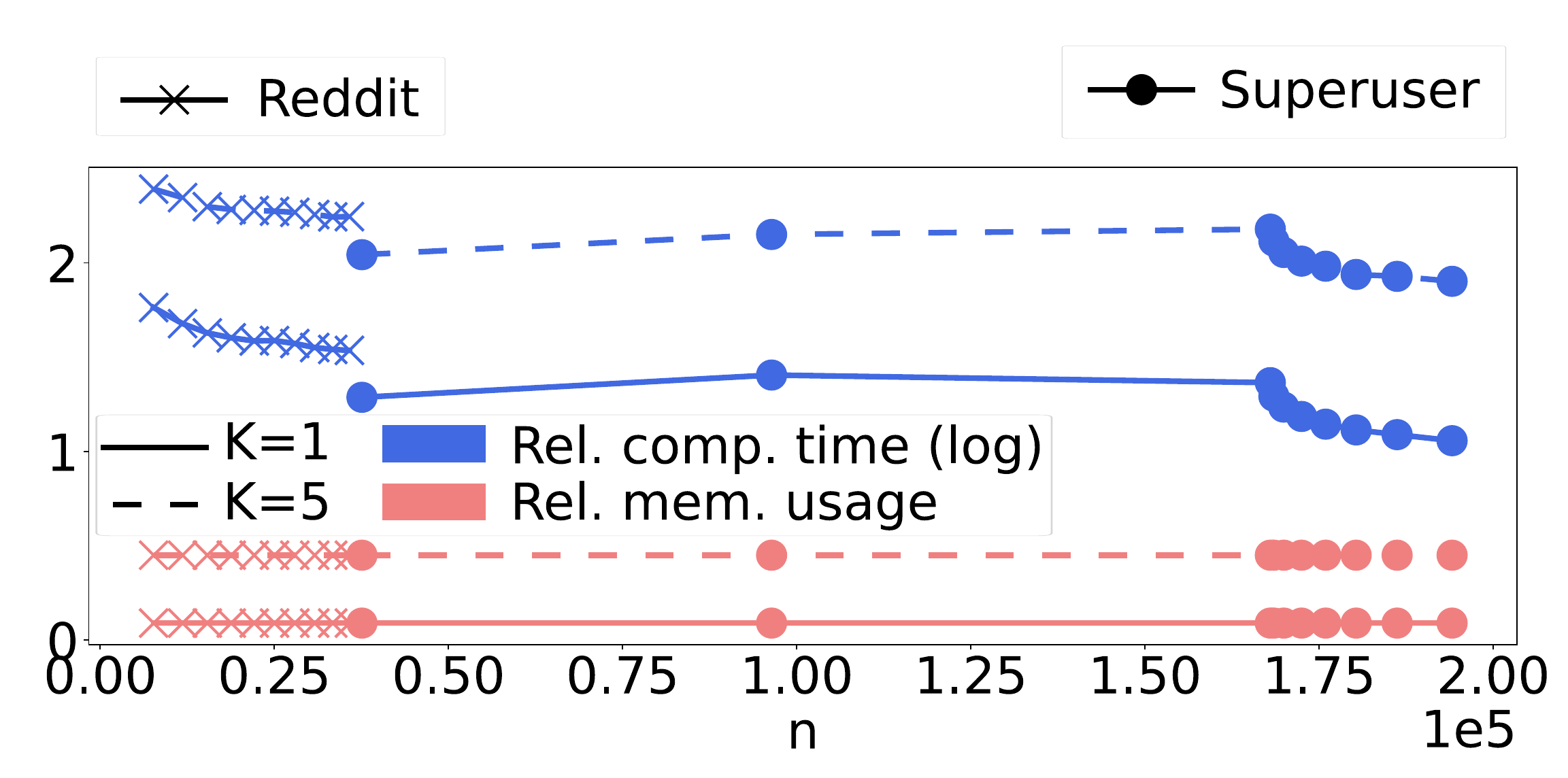}
    \caption{Relative computation time and relative memory usage for the Reddit and Superuser datasets to compute the distribution of the largest out-components thanks to the hashing framework compared to the matrix method depending on the number of nodes $n$.}
    \label{fig:rel_time}
\end{figure}

For the relative memory usage, the figure is blue meaning that we always gain memory. In fact, in this setup, only half the memory of the matrix method is required for the hashing framework. As for the Earth-Mover distance, there is a regime for small datasets where the accuracy is not satisfactory but for the large majority of the generated datasets, the hashing framework performs very well.

For the real datasets, we first show the results of the relative computation time and the relative memory usage of the hashing framework compared to the matrix method in Fig.~\ref{fig:rel_time}. Experimentally, we show that the hashing framework generally requires more time to compute the target distribution. Moreover, the relative computation time is linear with the number of hashing functions. That is, for $K=5$, that time is approximately $5$ times higher than for $K=1$ for both the Reddit dataset and the Superuser dataset. Overall, the general shape of the curves is in line with the results of generated data. For example, the relative computation time for the third point of the Reddit dataset, $n=15370$ and $m=85968$, is $198$, which coincide with the corresponding value of the generated datasets. Overall, the relative computation time decreases as the number of nodes increases, as expected.

Then, the memory required for the computation is linear with the number of hashing functions. We clearly see that, for $K=5$, the memory usage is $5$ times more than the one for $K=1$. The figures for real datasets are also in line with the figures for generated datasets. Obviously, with $K=1$ and $n_s = 0.3 \times n$, the computation requires less memory than with the full matrix algorithm for both real datasets by a factor $10$. But, more importantly, with $K=5$ and $n_s=0.3 \times n$, the hashing framework still requires only around $50\%$ of the memory of the matrix method.

\begin{figure}
    \centering
    \includegraphics[width=.65\textwidth]{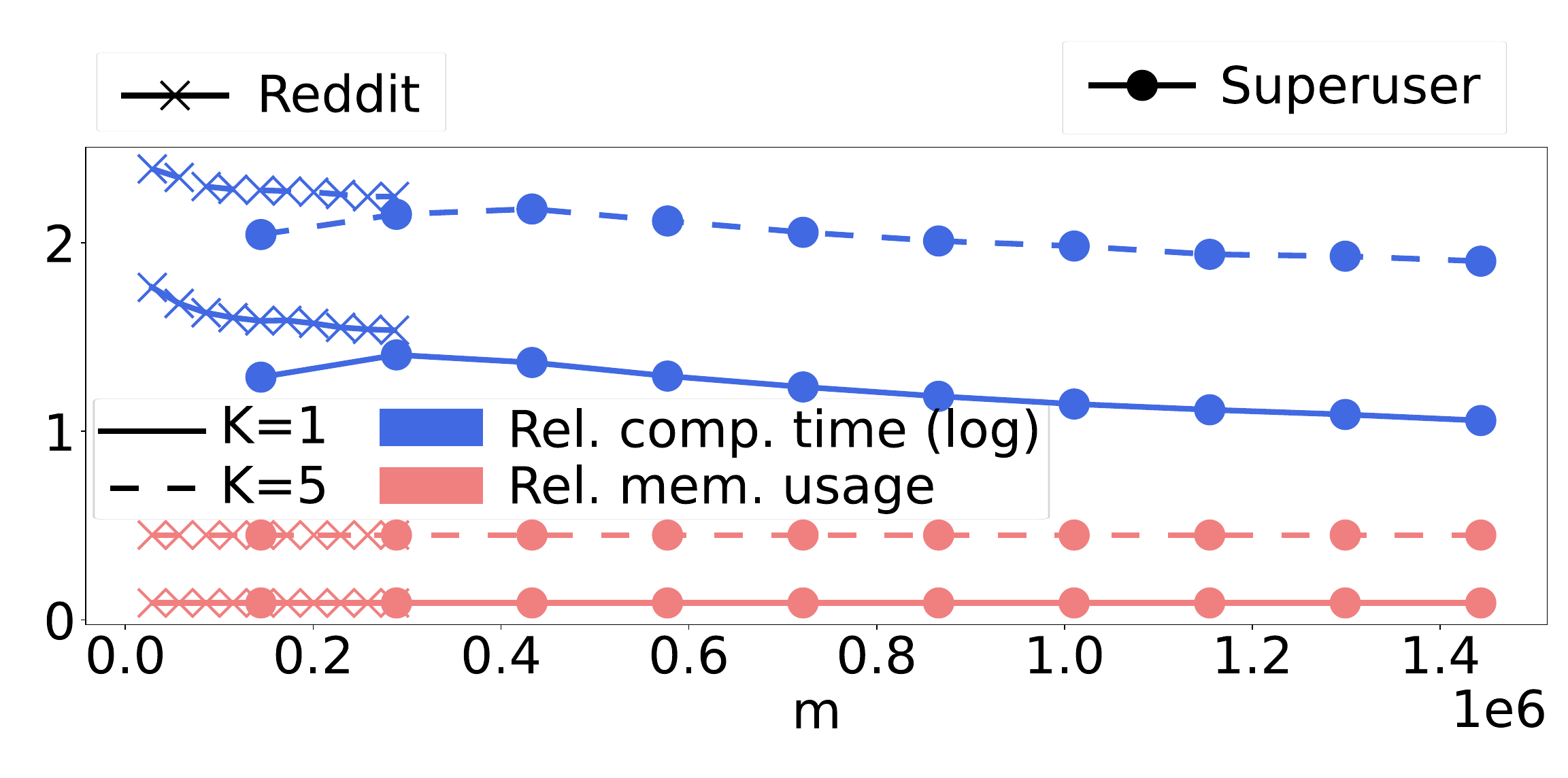}
    \caption{Relative computation time and relative memory usage for the Reddit and Superuser datasets to compute the distribution of the largest out-components thanks to the hashing framework compared to the matrix method depending on the number of events $m$.}
    \label{fig:mem_usage}
\end{figure}

\begin{figure}[!h]
\centering
\begin{minipage}{.5\textwidth}
  \centering
  \includegraphics[width=\linewidth]{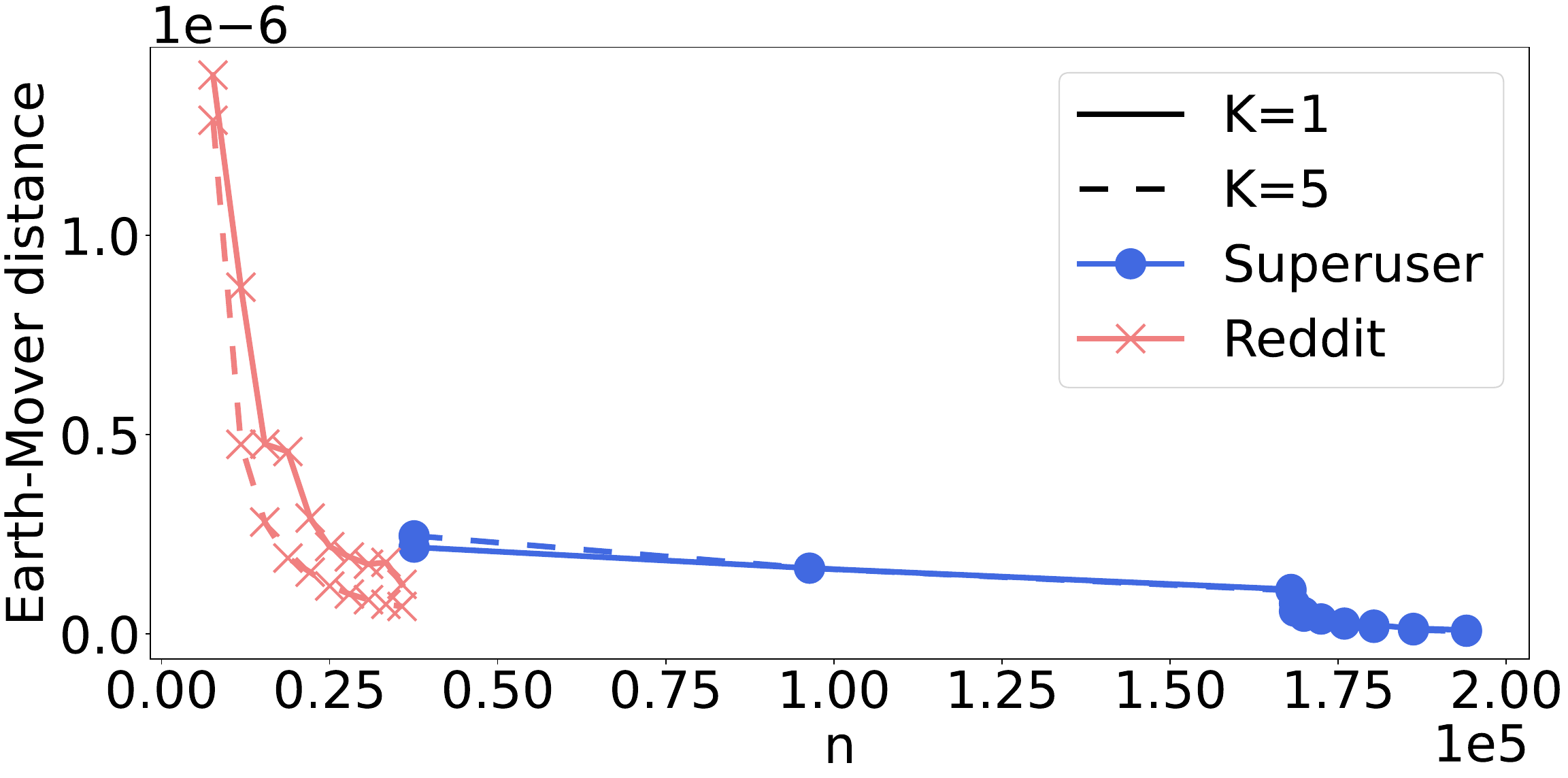}

\end{minipage}%
\begin{minipage}{.5\textwidth}
  \centering
  \includegraphics[width=\linewidth]{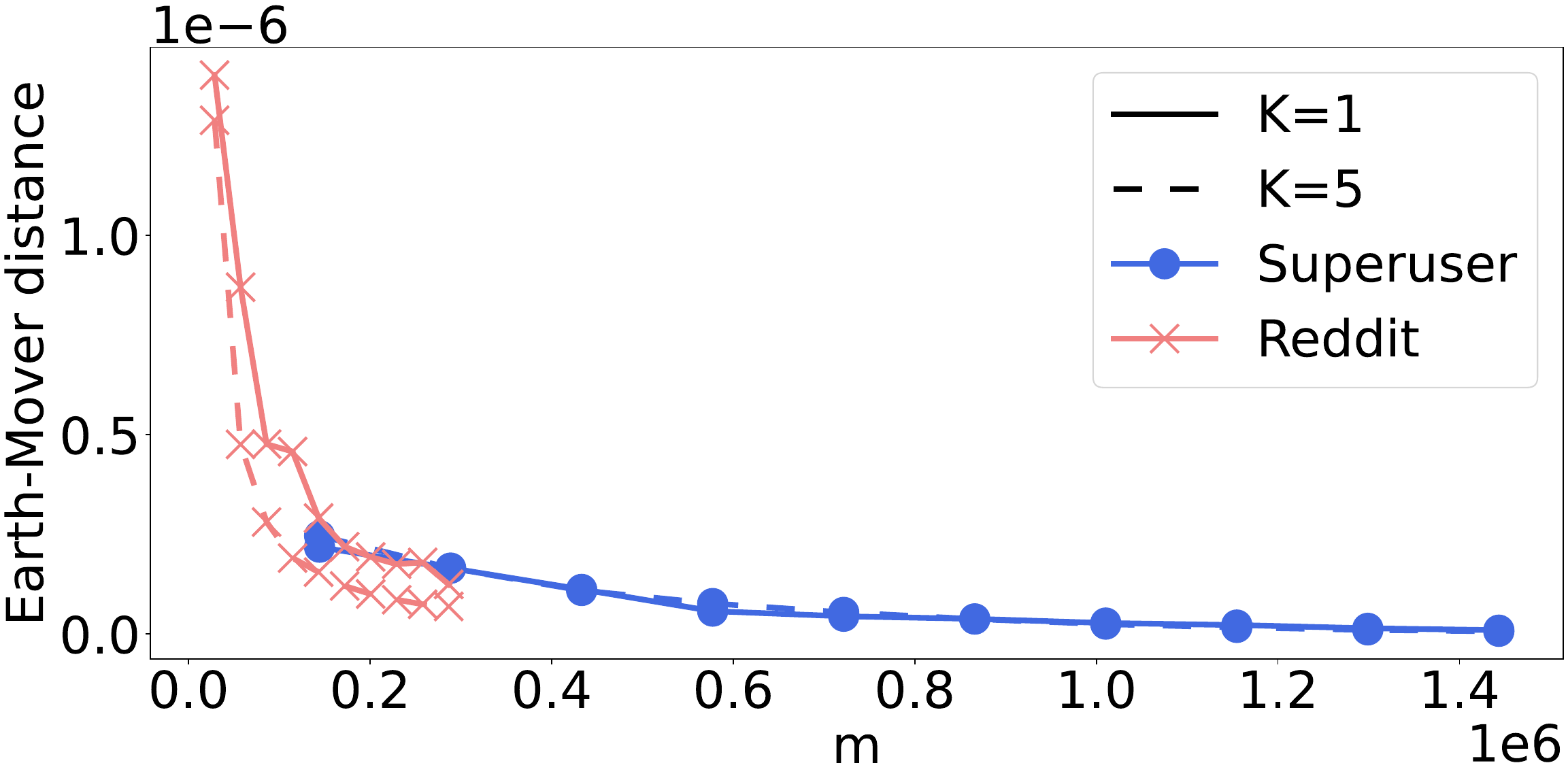}

\end{minipage}
\caption{Accuracy for the Reddit dataset and the Superuser dataset to compute the distribution of the largest out-components thanks to the hashing framework compared to the matrix method.}
\label{fig:accuracy}
\end{figure}

Finally, we report the accuracy of the hashing framework compared to the matrix method in terms of the Earth-Mover distance between the true distribution computed by the matrix method and the one estimated by the hashing method in Fig.~\ref{fig:accuracy}. Indeed, the quality of the results is important to assess the quality of the hashing framework. For both datasets, lower dimensions lead to lower accuracy of the framework. We see that the first few points, corresponding to networks of small sizes, have a significantly higher Earth-Mover distance (thus, lower accuracy) than the remaining ones. Overall, the shape of the curves still confirms the results with the generated datasets: the larger the network, the better the approximation. Secondly, as expected, the distance is lower for higher values of $K$. The accuracy of the method increases as there are more hashing functions.

Thus overall we can conclude that hashing is relevant in high dimension. There is a computation time gain for $m \geq 10^9$ in the generated datasets while memory usage remains lower and accuracy is good. Also, increasing the number of hashes leads to a linear increase in the memory usage and a linear increase in the computation time. Obviously, this increases the accuracy of the method. 

\section{Conclusion}

The continuous growth in the size of data bases requires new algorithms to process information. Moreover, structured data evolving over time represents an important challenge since it differs a lot from usual tabular data. To that end, we proposed a matrix algorithm that is able to compute both the out-components for every node of a temporal network and their sizes. Furthermore, to reduce the complexity of the analysis, we proposed a compression scheme based on hashing function that reduces the number of nodes of the network at the cost of some uncertainty. Uncertainty is lifted thanks to the use of several hashes in parallel. On each hashed graph, the matrix algorithm can be computed and, finally, all the information is merged to approximate the component matrix of the input network. Our framework is online and allows parallelization. Indeed, new nodes and news events can be processed as they come. Moreover, the different hashed graphs allow parallelization since they are independent. 
Additionally, hashing can make the computation private. If we do not observe the temporal network directly but only hashed versions of it and if hashes have some external randomness, our framework allows $\epsilon$-differential privacy \cite{Dwork2006}.
We believe that our work has a lot of potential applications. The first concrete user case is to use out-component sizes as the maximum number of nodes reachable during a spreading process. For example, it can be the maximum number of people infected by a virus from a single source. Or, on Twitter, it can be the maximum number of people a piece of news spread to. Secondly, our framework can be extended to other cases. In our work, we focused on out-components but we believe that many other quantities can be computed thanks to our compression scheme such as pairwise distances between nodes. Also, we believe that the hashing framework can be rewritten with an algebraic formulation. This would open up the work to linear problems and linear solvers. In fact, the reconstruction of the matrix could be tackled in many different ways making the framework more flexible. Moreover, privacy preserving algorithms are particularly interesting for security or privacy reasons. The work we propose can efficiently make algorithms on temporal networks private. Indeed, adding randomness in the data can lead to prevent the identification of the source of the data.
Most importantly, our hashing framework transforms a temporal network into a series of smaller datasets that can be used to infer properties of the initial dataset without direct access to it. This can be very beneficial in the processing of sensitive information.

\section*{Acknowledgement}
This work has been supported by the DATAREDUX project, ANR19-CE46-0008. MK was supported by the CHIST-ERA project SAI: FWF I 5205-N; the SoBigData++ H2020-871042; the EMOMAP CIVICA projects; and the National Laboratory for Health Security, Alfréd Rényi Institute, RRF-2.3.1-21-2022-00006. PB was supported by the CHIST-ERA-19-XAI-006, for the GRAPHNEX ANR-21-CHR4-0009 project.

%
%
%

\end{document}